\documentclass[sigconf, nonacm]{acmart}

\newcommand\vldbdoi{XX.XX/XXX.XX}
\newcommand\vldbpages{XXX-XXX}
\newcommand\vldbvolume{14}
\newcommand\vldbissue{1}
\newcommand\vldbyear{2023}

\newcommand\vldbavailabilityurl{URL_TO_YOUR_ARTIFACTS}
\newcommand\vldbpagestyle{plain} 

\usepackage{amsmath}

\usepackage{amssymb}
\usepackage{amsfonts, bm}
\usepackage{algorithmic}
\usepackage{graphicx}
\usepackage{hyperref}
\usepackage{subfigure}
\usepackage{blindtext}
\usepackage{balance}
\usepackage{textcomp}
\usepackage{xcolor}
\usepackage[export]{adjustbox}
\usepackage{enumerate}
\usepackage{booktabs}
\usepackage{arydshln}
\usepackage{multirow}
\usepackage[misc]{ifsym}
\usepackage{listings}
\usepackage{amsmath}
\begin{document}
\title{NHtapDB: Native~HTAP~Databases}


\begin{abstract}
\textbf{H}ybrid \textbf{T}ransactional/\textbf{A}nalytical \textbf{P}rocessing (HTAP) database in a natural productive environment must leverage multimodal data to generate valuable real-time business insights, and execute OLAP queries in-between online transactions. State-of-the-art and state-of-the-practice systems adopt (1) a separate database and disparate business applications that leverage machine learning techniques to generate real-time business insights and (2) dual-format stores to guarantee the performance of different workloads---row-based storage for OLTP workloads and column-based storage for OLAP workloads. They fail to achieve the above goals because of massive data transfer overhead rooted in separate systems and dual-format stores.
To this end, we propose NHtapDB, the first native HTAP database, providing business insight in real-time (within milliseconds to seconds). NHtapDB (1) provides a near-data machine learning framework to facilitate generating real-time business insight, and predefined change thresholds will trigger online training and deployment of new models, and (2) offers a mixed-format store to guarantee the performance of HTAP workloads, especially the hybrid workloads that consist of OLAP queries in-between online transactions.
We make rigorous test plans for NHtapDB with an enhanced state-of-the-art HTAP benchmark.

\end{abstract}
\maketitle
\pagestyle{\vldbpagestyle}
\begingroup\small\noindent\raggedright\textbf{PVLDB Reference Format:}\\
Guoxin Kang, Lei Wang, Simin Chen, Jianfeng Zhan. NHtapDB: Native HTAP Databases. PVLDB, 
\vldbvolume(\vldbissue): \vldbpages, \vldbyear.\\
\href{https://doi.org/\vldbdoi}{doi:\vldbdoi}
\endgroup
\begingroup
\renewcommand\thefootnote{}\footnote{\noindent
\Letter~Jianfeng Zhan is the corresponding author.

\noindent 
This work is licensed under the Creative Commons BY-NC-ND 4.0 International License. Visit \url{https://creativecommons.org/licenses/by-nc-nd/4.0/} to view a copy of this license. For any use beyond those covered by this license, obtain permission by emailing \href{mailto:info@vldb.org}{info@vldb.org}. Copyright is held by the owner/author(s). Publication rights licensed to the VLDB Endowment. \\
\raggedright Proceedings of the VLDB Endowment, Vol. \vldbvolume, No. \vldbissue\ %
ISSN 2150-8097. \\
\href{https://doi.org/\vldbdoi}{doi:\vldbdoi} \\
}\addtocounter{footnote}{-1}\endgroup

\ifdefempty{\vldbavailabilityurl}{}{
\vspace{.3cm}
\begingroup\small\noindent\raggedright\textbf{PVLDB Artifact Availability:}\\
NHtapDB is a vision paper, so it is not available for lacking of experiments.
\endgroup
}

\begin{figure}[!t]
 \centering
 \includegraphics[scale=0.28]{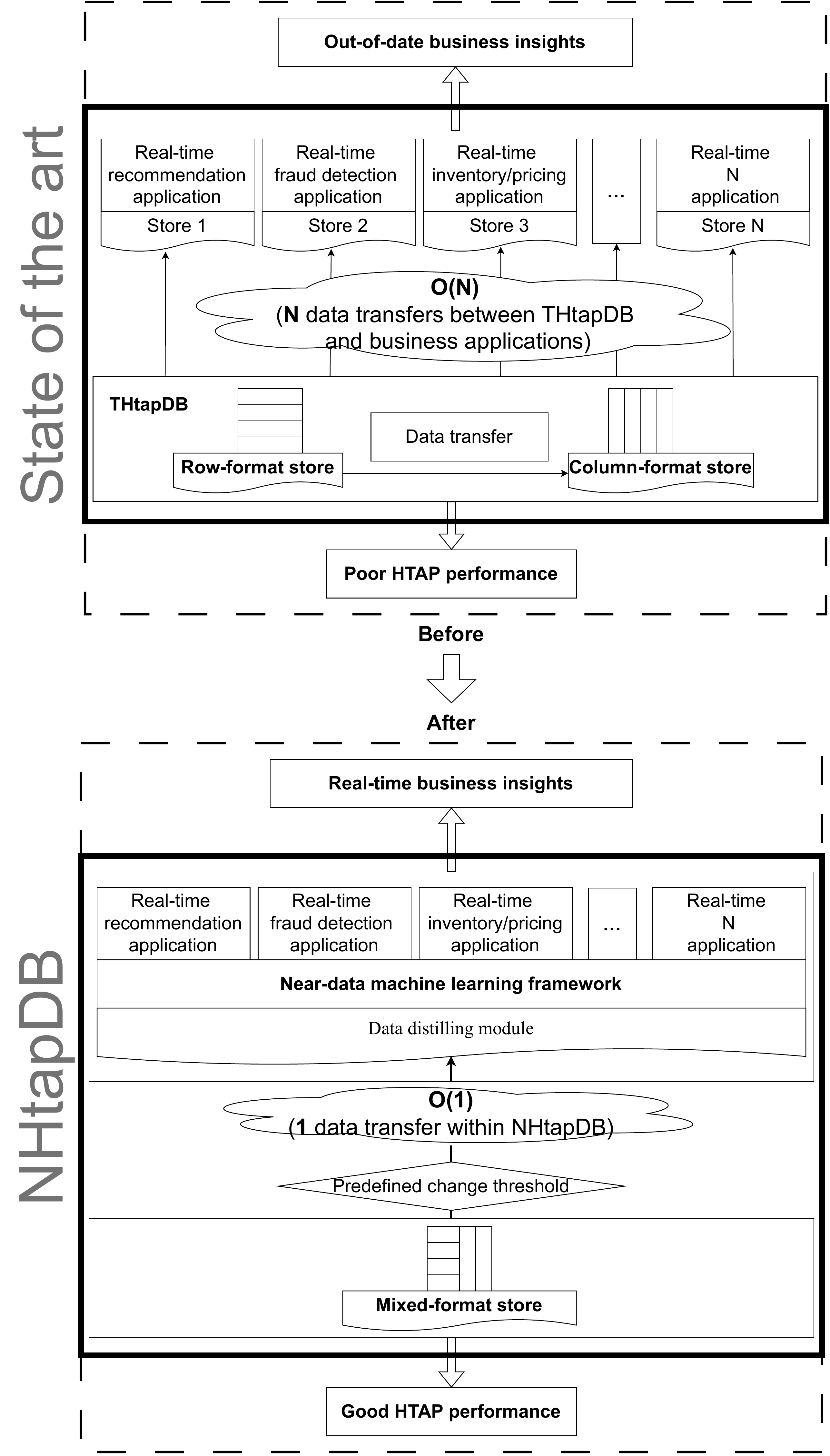}
 \caption{The motivation for the Native HTAP database: massive data transfer overhead results in the lack of capability of generating real-time business insights:  the state-of-the-art systems (before) vs. NHtapDB (after).}
 \label{fig: 1-0}
 \vspace{-0.45cm}
\end{figure}

\section{Introduction}
The goal of the traditional HTAP database (THtapDB) is to support Online Analytical Processing (OLAP) on the fresh data generated by the Online Transaction Processing (OLTP)~\cite{lahiri2015oracle, larson2015real, lee2017parallel, 9835628}.
OLTP workloads generally read and write a small number of rows by index.
OLAP workloads are read-intensive and involve complex queries on a few columns but numerous rows.
Despite the abundance of THtapDB, they perform poorly in generating real-time business insights and guaranteeing the performance for HTAP workloads because of the vast data transfer overhead rooted in (1) a separate database, disparate business applications leveraging machine learning techniques,  and (2) dual-format stores: row-based storage for OLTP workloads and column-based storage for OLAP workloads.

First,  state-of-the-art or state-of-the-practice THtapDB workloads often run for a relatively long time (from minutes to hours), incapable of providing real-time business insights and preventing it from satisfying the fleeting needs of customers.
Consistent with the previous work~\cite{9835647}, the implication of real-time emphasizes performing a task like data analysis or user behavior simulation interactively within milliseconds to seconds.
THtapDB deployer focuses much of its effort on guaranteeing online transactions and complex query performance.
They ignore the enormous benefits that real-time business data can bring to business applications~\cite{thennakoon2019real, pereira2019online} and, as a result, fail to build the necessary infrastructure.

As shown in Figure~\ref{fig: 1-0}, state-of-the-art architecture is separated into two parts: THtapDB and disparate business applications.
Self-governed business applications leverage the data loader~\cite{ofeidis2022overview} to retrieve business data from the database, pre-process the business data into training samples, and then feed the training samples to the business applications. As an independent OS process, each self-governed business application needs its own data loader instance to transfer data from databases. However, the long distance between THtapDB and business applications results in a long data turnaround time and O(N) data transfer, as shown in Figure~\ref{fig: 1-0}, making it challenging to satisfy the fleeting needs of customers.  

 THtapDB stores a large amount of up-to-date data critical for business applications, but there is an urgent need to establish enabling facilities to generate real-time business insights  from sharing business data.
In E-commerce, many self-governed business applications, such as real-time recommendations, fraud detection, inventory/pricing, etc., need multiple real-time customer services based on the same business data.
A typical scenario is real-time business insights in e-commerce that capture the real-time preferences of the customer to generate more accurate recommendations (predictions), which is a significant source of revenue for business applications. For example, in Amazon, the real-time recommendation can increase transactions by 35\%~\cite{zhou2022competing}.
If the recommendation is not real-time, the customer will likely leave the current session and purchase in another e-commerce application.

Besides, state-of-the-art THtapDB does not utilize multimodal data very well.
Multimodal data refers to the different data types from different sources that may be complementary~\cite{oviatt2018handbook}.
Multimodal data includes more details about customer behaviors that happened in business applications, such as customer clicks and customer reviews in e-commerce platforms, such as Taobao~\cite{shi2019virtual}, so the fusion of these multimodal data could provide more accurate real-time business insights.

Second, the actual HTAP workloads consist of hybrid transactions that execute OLAP queries in-between online transactions~\cite{9835647, li2022htap}, rather than the separate OLTP and OLAP workloads. However, state-of-the-art or state-of-the-practice THtapDB adopts row-format, column-format, or dual-format stores that exaggerate the  challenges in guaranteeing the HTAP workload performance. 

For example, in e-commerce,  customers prefer to purchase the best-selling commodity within their budget and other constraints, which is a typical HTAP workload. The related statements in the HTAP transaction are as follows. 
\begin{lstlisting}[language=SQL]
1) SELECT MAX(ws_quantity) 
   FROM web_sales
   WHERE ws_price between 64 and 64 + 16; 

2) UPDATE customer SET c_balance = 1024
   WHERE c_id = 256;  
\end{lstlisting}
The best-selling commodity selection calls the OLAP MAX function (ws\_quantity) in structured query language (SQL), which is performed between online transactions  --- purchase operations involving banking and credit card activity (c\_balance).

Row-based storage allows online transactions to locate the rows to update quickly. At the same time, column-based storage permits read-intensive OLAP queries to scan and aggregate operations over specific columns quickly.
Nevertheless, a single row-based or column-based data organization is insufficient for HTAP workloads that have OLAP workloads in-between OLTP operations.

Using a dual-format store~\cite{lahiri2015oracle, huang2020tidb} will introduce data transfer overhead between the row-based store and column-based store as shown in Figure~\ref{fig: 1-0}. 
To keep data fresh, row-based data updates must be propagated to column-based storage as soon as possible.  The previous work~\cite{9835647} revealed the peak HTAP performance using a dual-format store is  lower than the peak OLTP performance by one order of magnitude.
Consequently, a more suitable data organization for HTAP workloads is one of THtapDB's reform goals.

\textbf{Our solution.} As shown in Figure~\ref{fig: 1-0}, to mitigate the heavy data transfer overhead O(N) between THtapDB and business applications, we built the native HTAP databases (NHtapDB), which not only makes a near-data machine learning infrastructure for acquiring real-time business insights but also handle HTAP workloads well.
 To provide real-time business insights, NHtapDB adopts the near-data machine learning framework that leverages predefined change thresholds to trigger online training to update the models in real-time. We implement data distilling instance that loads the real-time data, pre-processes it into training samples, and feeds training samples into models. And the near-data machine learning framework implements unified management for multiple models.
When NHtapDB receives customer queries, the corresponding model must be updated until the customer leaves to satisfy their needs. The recommendation model must be updated in real-time to recommend the highly matched commodities lists if a customer searches for a commodity. As soon as a customer purchases a product, the real-time fraud detection model must be updated to identify fraudulent and anomalous activity associated with the transaction. NHtapDB implements a mixed-format store, which eliminates the inherent data transfer overhead between row-based and column-based stores that exist in the dual-format store.
With the cooperation of the near-data machine learning framework, the mixed-format store, and other infrastructure, the data transfer overhead is reduced from N+1 times to 1 time. A simple upper-bound performance model in Section~\ref{motivation} shows the gap between the data transfer overhead of the two systems using near-data machine learning framework (NHtapDB) and separate machine learning systems (THtapDB) is 10,000 times in the case of 50 business applications.  

Overall, the spotlights of NHtapDB can be summarized as follows:

(1) For real-time business insights,  NHtapDB integrates a near-data machine learning framework to mitigate data transfer overhead between the database and the business applications.
The machine learning engine abstracts three essential elements: state, action, and reward. 
To dynamically satisfy the customer's preferences, the machine learning engine uses the current state of the customer session as a new training sample to immediately update the model, then outputs the action, and finally calculates the reward to evaluate the action.

(2) NHtapDB provides a mixed-format store to handle the HTAP workloads with zero data transfer overhead. 
Mixed-format store splits the records into row-based update partitions and column-based read-only partitions.
The row-based partitions are responsible for update operations, and the column-based partitions are leveraged for analytical queries.
Mixed-format store eliminates the inherent data transfer overhead in dual-format store.
Besides, the mixed-format store adopts the split write-ahead logging (WAL)~\cite{mohan1992aries} to guarantee insert and delete performance.

To demonstrate NHtapDB, we make rigorous demonstration plans.
First, we will demonstrate the near-data machine learning framework provides real-time business insights. Second, we will demonstrate the mixed-format store can handle the HTAP workload well.

\begin{figure}[!t]
 \flushleft
 \includegraphics[scale=0.43]{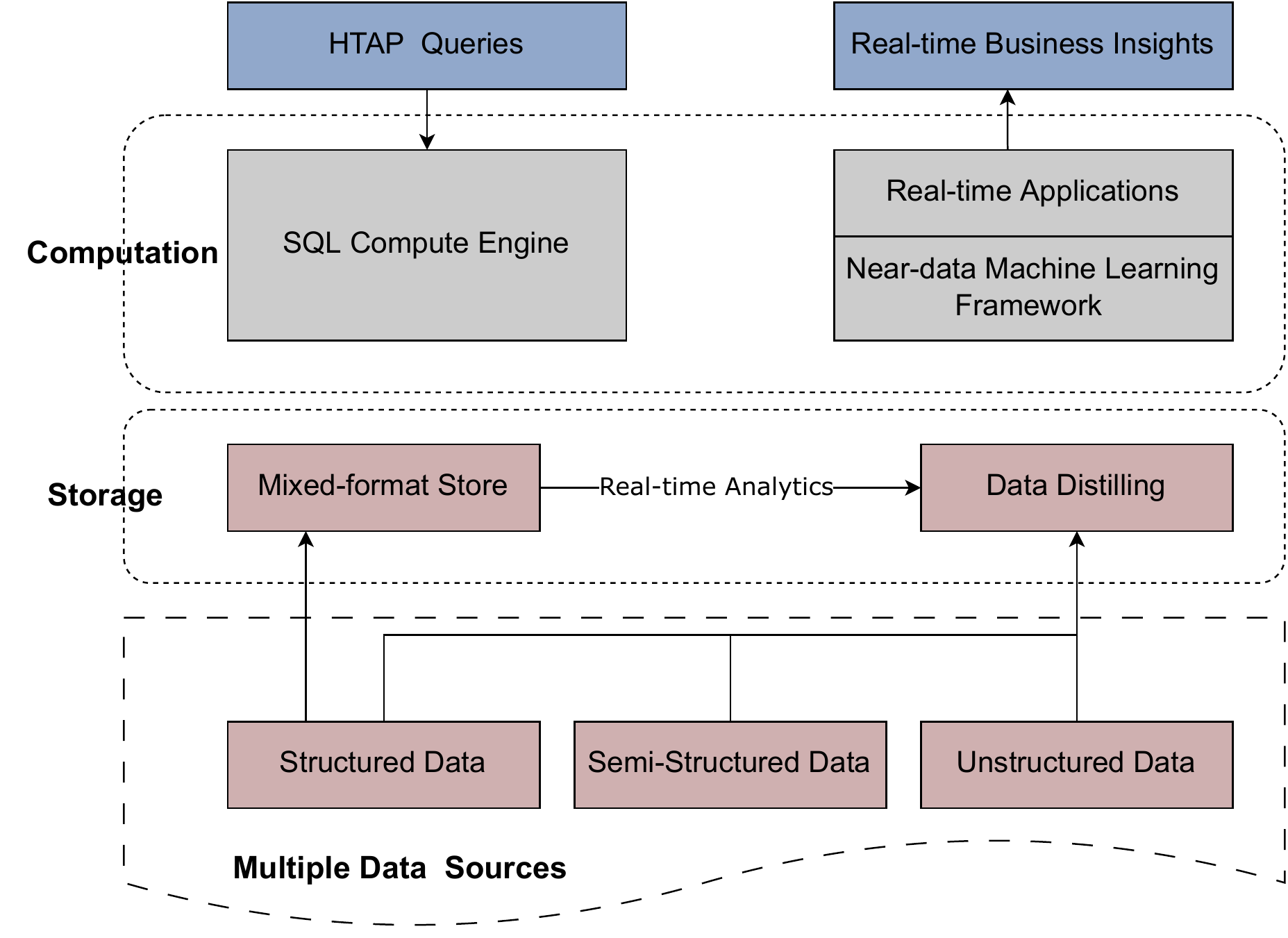}
 \caption{ The overview of NHtapDB.}
 \label{fig: 3-0}
\end{figure}

\section{Why do we need NHtapDB?}~\label{motivation}

To demonstrate the motivation, we build a simple upper-bound performance analysis model to evaluate the application (end-to-end) latencies under THtapDB and NHtapDB. We mainly consider the near-data machine learning framework's effect in NHTapDB. While under THtapDB,  as an independent OS process, self-governed business applications leverage the data loader~\cite{ofeidis2022overview} to retrieve business data from the database and provide the service. This simple model fails to consider the effect of different stores, e.g., mixed-format stores under NHTapDB vs. dual-formal or single-store in THtapDB.

The application's latency includes two components:  the data transfer latency and  the computation latency. 
Data transfer latency is the key to significantly reducing overall latency.
Consequently, we compare the data transfer latency between THtapDB and NHtapDB.
We assume there are 50 business applications (N=50), and each business application needs to deal with the 1GB data analysis.
For the THtapDB solution, the total transfer bandwidth is 500 MB/Second (using the state-of-art NFS solution for data transfer), and that of NHtapDB is 100GB/Second (using the memory space access on the same OS process).
The data transfer latency of the THtapDB solution is 100 seconds (the transfer bandwidth for each application is 10MB/S, and the transfer latency of each application is 100 seconds).
At the same time, that of NHtapDB is only 0.01 seconds (the transfer bandwidth is constant at 100GB/Second, and the transfer latency of each application is 0.01 seconds). The gap between the data transfer overhead of the two systems  is 10,000 times. 

\section{NHtapDB Overview}
NHtapDB is a powerful HTAP database that ensures HTAP performance and provides real-time business insights based on multimodal data fusion.
In this section, we sketch one possible architecture for NHtapDB, depicted in Figure~\ref{fig: 3-0}.
NHtapDB is composed of two main layers: the computation layer and the storage layer.

\subsection{Computation layer}
An efficient computation layer is supported by two modules:

(1) Near-data machine learning engine aims at accelerating business data into insights in real-time to satisfy the fleeting needs of customers.
It defines and implements three essential elements to support online training -- state, action, and reward.
ML engine first receives the current state of the customer, then performs the suitable action, and finally calculates the reward to evaluate the action.
ML engine needs to reuse other state-of-the-art techniques, such as feature engineering.

(2) The SQL compute engine is stateless, scalable, and aware of storage.
It chooses the cheapest logical plan according to estimated execution costs and then transforms the logical plan into a physical execution plan based on the storage layout.

\subsection{Storage layer}
The storage layer comprises a mixed-format store and a data-distilling module.

Mixed-format store guarantees HTAP performance by eliminating the data update propagation delays inherent in dual-format stores.
In dual-format stores, the OLAP delays have two sources, one is the data update propagation delays from row-based store to column-based store, and the other is the OLAP execution delays.
Mixed-format store splits row records into row-based update partitions and column-based non-update partitions so that online transactions and analytical queries execute in parallel.
To improve SQL performance, a mixed-format store needs cooperation with other techniques, such as caching, indexes, and SQL computation engines.

The data distilling module adopts a novel lightweight multimodal data fusion technique to acquire real-time business insights.
When the predefined change thresholds are triggered, it will receive structured business data from the database.
In addition, it receives multimodal business data from multiple data sources, such as stream systems.
Following data cleansing, feature extraction, and other processes, the above business data is formed into training samples, which are then fed into the computation layer for online learning.

\section{NHtapDB Design}
In this section, we introduce two key design aspects of NHtapDB depicted in Figure~\ref{fig: 3-1}.
We describe the use case of near-data machine learning framework and mixed-format store in detail below.

\begin{figure}[!t]
 \centering
 \includegraphics[scale=0.8]{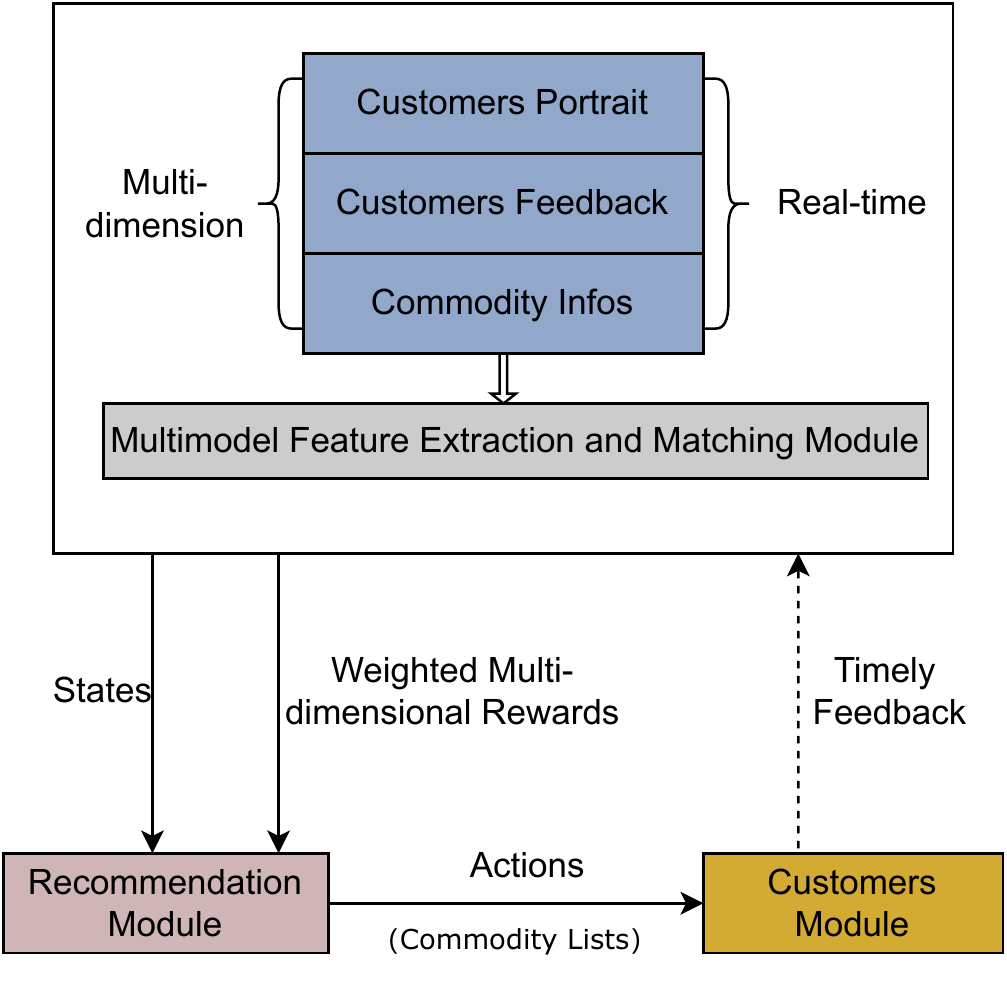}
 \caption{Machine learning model instance --- real-time recommendation model.}
 \label{fig: 3-2}
\end{figure}

\subsection{Near-data machine learning framework}
Near-data machine learning framework defines and implements three essential elements on top of heterogeneous models for real-time business insights.
This subsection explains the three elements and the machine learning model instance as shown in Figure~\ref{fig: 3-2}.

\subsubsection{Essential elements.}

Near-data machine learning framework defines and implements three elements: state S, action A, and reward function R.

\emph{State S} is the set of all possible states, ${S^t}$ represents the state at time step t. 

\emph{Action A}  is the set of available actions depend on states, ${A^t}$ is the action taken at time step t.

\emph{Rewards R} provides a reward to assess the actions selected.

\subsubsection{Machine learning model instance}
We give the instance of a real-time recommendation model, which reuses the  three essential

\begin{table}[ht]
    \caption{Partial features of multi-dimension data ($p_1$--$p_2$: customer portrait features; $c_1$--$c_5$: click feedback features; $q_1$--$q_2$: text/image query feedback features; $r_1$--$r_2$: additional labels feedback features; $i_1$--$i_3$: commodity information features).}
    \label{tab:3-2}
    \scalebox{0.85}{
\begin{tabular}{cccc}
	\toprule
		\textbf{Variate}  & \textbf{Features}   & \textbf{Description} & \textbf{Description} \\
  \midrule
        $p_1$  & time    &  the time of day     & string     \\
		$p_2$  & location    & the location  of customer   & string     \\
  \midrule
		$c_1$  & pv    &  commodity page view     & bool     \\
		$c_2$  & buy    & buy commodity     & bool     \\
		$c_3$  & cart    & add commodity to  shopping cart     & bool     \\
		$c_4$  & favorite    & favorite commodity     & bool    \\
		$c_5$  & duration    & commodity page view duration     & string    \\
    \hdashline
		$q_1$  &  text query   &  query in  natural language    & string    \\
		$q_2$  & image query    & query in image    & string    \\
     \hdashline[1pt/5pt]
		$r_1$  & price    & real-time price range    & float    \\
		$r_2$  & inventory    & real-time inventory quantity     & int    \\
  \midrule
        $i_1$ &  category & commodity category & one-hot   \\
        $i_2$ &  subcategory & commodity subcategory & one-hot   \\
        $i_3$ &  style & commodity style & string \\
	\bottomrule
\end{tabular}
}
\end{table}
\noindent elements and defines recommendation, customer, and  multimodal feature extraction modules, as explained below.

At the time step $t$, the recommendation module precepts the current state $S^t$ of the customers and recommends the suitable commodity list to the customer. Next, it receives weighted multi-dimensional rewards $R^t$ representing the quality of the recommendation, and then it receives the new state $S^{t+1}$ at time step $t$+$1$.

\emph{Customers module.}\label{sec:3.2.1}
Customers send timely feedback and receive the recommended commodity list.
Timely feedback reveals interactions between customer and commodity in the current state, including click feedback and search feedback.
Click feedback implied real-time customer preference information, which contains commodity page view (pv), adding commodity to shopping cart (cart), buying the 
commodity (buy), favorite commodity (favorite), and commodity browsing duration (duration).
And search feedback refers to the real-time text/image query and additional labels, directly indicating the customer's needs.
For example, in Taobao~\cite{shi2019virtual}, customers could use text or images to search for the commodity. 
They could also use additional labels to filter targeted commodities quickly, and common labels include real-time inventory quantity and price range.
 Above feedback is sent to the multimodal feature extraction and matching module for further processing.

\emph{Multimodal feature extraction and matching module } 
It collects multi-dimension data for feature extraction and matching, which includes customer portraits, customer feedback, and commodity information. 
And the feature is used for the current state 
 $S^t$ representation.
Features of the above data are depicted in table~\ref{tab:3-2}.
Customer portrait features refer to the time of day, location, customer scenario, etc. 
Given a customer $i$, the corresponding portrait is mapped to feature $p_i^t$.
Customer feedback is detailedly described in the customer module.
The click feedback feature is denoted as $c_i^t$.
Text query feature is denoted as  $q_{text}^t$.
And the image query feature is denoted as $q_{image}^t$.
Real-time analytics generates an additional label that is denoted as $r_i^t$.
Commodity information collects the basic attributes of a commodity, such as a commodity category, commodity subcategory, and commodity style, which is denoted as $i_i^t$.

Next, the current state $S^t$ is delivered to the recommendation module.

\emph{Recommendation module.}
It receives the current state $S^t$ and recommends the commodity list to the customer.
Later, it receives weighted multi-dimensional rewards $R^t$, which evaluate the quality of the recommendation.
Multi-dimensional rewards have six parts: customer portrait reward $R_p^t$, click feedback reward $R_c^t$, text query reward $R_{text}^t$, image query reward $R_{image}^t$, additional label reward $R_r^t$, and commodity information reward $R_i^t$.
And the final reward $R^t$ at the time step t is:
\begin{equation}
 R^t = \beta + \lambda_1 R_p^t + \lambda_2 R_c^t + \lambda_3 R_{text}^t + \lambda_4 R_{image}^t + \lambda_5 R_r^t + \lambda_6 R_i^t  \label{XX}
\end{equation}

In turn, the recommendation module proceeds with the next round of interaction with the customer module at the time step $t+1$.

\begin{figure}[!t]
 \centering
 \includegraphics[scale=0.54]{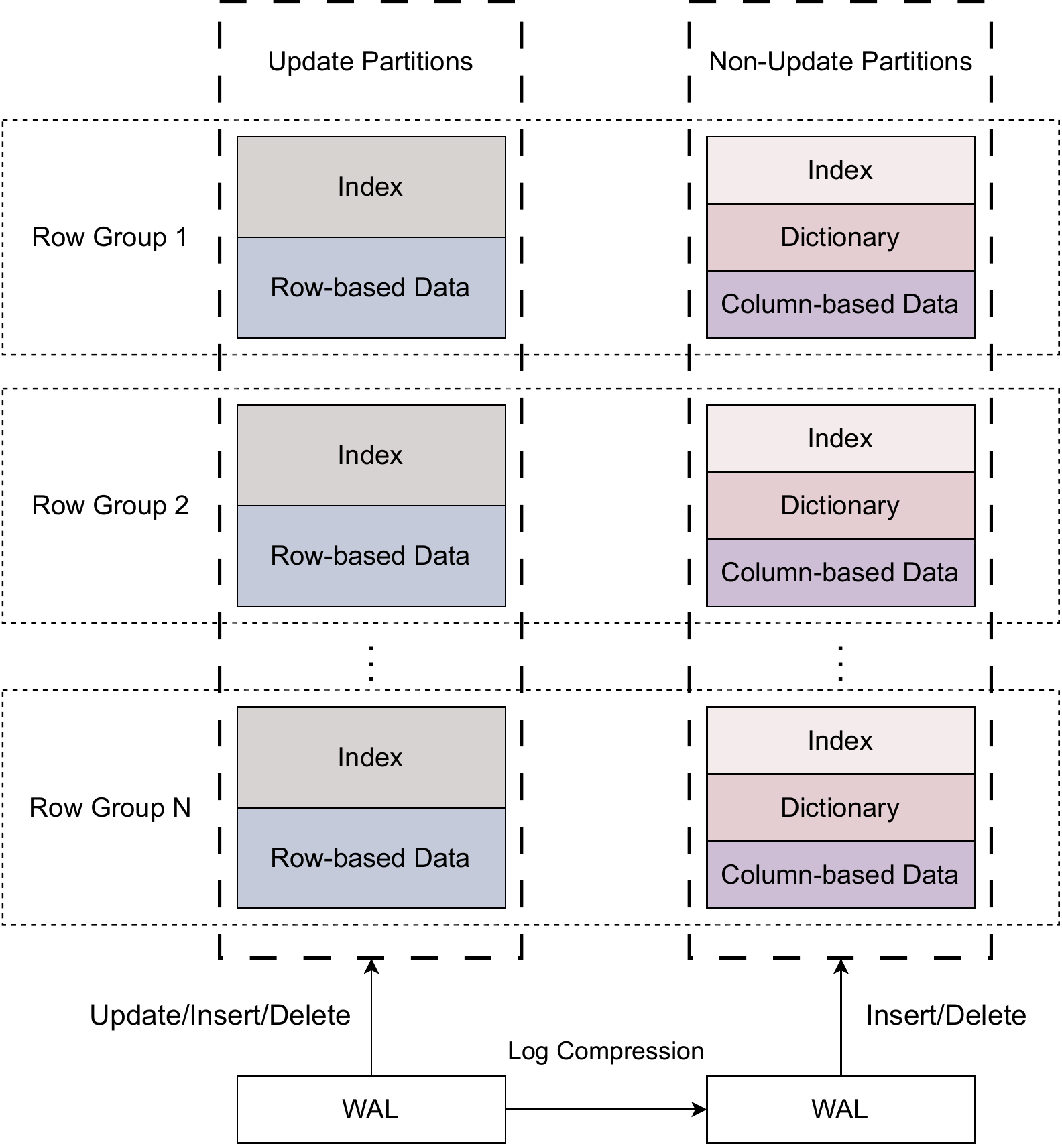}
 \caption{Mixed-format store architecture.}
 \label{fig: 3-1}
\end{figure}

\subsection{Mixed-format Store}
NHtapDB adopts a mixed-format store to guarantee the performance of HTAP workloads.
Mixed-format store architecture is shown in Figure~\ref{fig: 3-1}.
Mixed-format store adopts the range partition strategy to divide all row records into  $row$  $groups$ evenly to leverage modern computers' multi-core parallelism.
The $row$ $groups$ splits row records into $update$ 
 $partitions$ and $non$-$update$ $partitions$ based on whether their column attributes are updated.
The $update$ $partitions$ are placed in row-based format for OLTP performance, and the $non$-$update$ partitions are placed in column-based format for OLAP performance.
For example, the CUSTOMER table in the TPC-C benchmark places the $C\_ID$, $C\_BALANCE$, and $C\_DATA$ attributes into the row-based store, and places the other attributes into the column-based store.
At the same time, the mixed-format store avoids the data update propagating overhead between row-based and column-based data.
The mixed-format store must cooperate with state-of-the-art indexes and dictionary techniques~\cite{luo2019umzi, riegger2019mv, boissier2018hybrid}  to improve SQL performance.

NHtapDB adopts split write-ahead logging (WAL)~\cite{mohan1992aries} for atomicity and durability.
WAL sequentially records the transactions' prewritten, commit, and rollback behaviors into the persistent devices.  
There are three kinds of log items -- update log items, insert log items and delete log items.
The data in update partitions are updated as update log items.
The original insert and delete log item is split into a row log item and a column log item.
Only when the row log item is committed the corresponding column data are inserted or deleted as the column log item.
We adopt a log compression strategy to ease the unnecessary insert and delete pressure of column-based data.
Log compression strategy deletes column log items whose row log entries are rollback. The original log item will not be committed until both the row and column log items have been committed.

\section{Demonstration}

NHtapDB connects the underlying storage and the upper machine learning models, shortens the data turnaround time, and uses real-time analytics to help provide real-time business insights. 

Our demonstration plan aims to show that.

\begin{enumerate}[a)]
\item \textbf{Near-data machine learning framework} provides real-time business insights with low data transfer overhead.
\item \textbf{Mixed-format store} guarantees HTAP performance.
\end{enumerate}

For the first objective, we conduct experiments on the multimodal dataset~\cite{InterBERT}. Near-data machine learning framework has one test case.
\begin{enumerate}[1)]
\item $Test$ $case$ $1:$ Evaluating the data transfer overhead between database and business applications.
\end{enumerate}

For the second objective, 
we use the OLxPBench suite~\cite{9835647}, which is the first benchmark addressing the necessity of introducing HTAP workloads that execute OLAP queries in-between online transactions~\cite{9835647, li2022htap} and providing multiple HTAP workloads.
We ran the OLxPBench in different configurations by varying the type and sending rate of workload. In the demonstration, we compare NHtapDB with state-of-the-practice HTAP databases -- TiDB~\cite{huang2020tidb}.

\begin{enumerate}[2)]
\item $Test$ $case$ $2:$ Comparing the HTAP performance of TiDB and NHtapDB.
\end{enumerate}





\bibliographystyle{ACM-Reference-Format}
\bibliography{sample}

\end{document}